\documentclass[9pt,conference]{IEEEtran}
\usepackage{amssymb,amsthm,amsmath,array}
\usepackage{graphicx}
\usepackage[caption=false,font=footnotesize]{subfig}
\usepackage{xspace}
\usepackage[sort&compress, numbers]{natbib}
\usepackage{stmaryrd}
\usepackage{xcolor}
\usepackage{mathtools}
\usepackage{float}
\usepackage{textcomp}
\usepackage{glossaries}

\newacronym{VSM}{VSM}{vital sign monitoring}
\newacronym{RBM}{RBM}{random body movement}
\newacronym{DSP}{DSP}{digital signal processing}
\newacronym{US}{US}{Unlimited Sampling}
\newacronym{SINR}{SINR}{signal-to-interference-plus-noise ratio}
\newacronym{NICU}{NICU}{Neonatal Intensive Care Unit}
\newacronym{CPI}{CPI}{coherent processing interval}
\newacronym{SNR}{SNR}{signal-to-noise ratio}
\newacronym{ADC}{ADC}{analog-to-digital converter}

\begin{document}
\title{Recent Developments in Contactless Monitoring \\ Vital Signs Using Radar Devices}
\author{\IEEEauthorblockN{
        Gabriel Beltrão\IEEEauthorrefmark{2} 
        and Udo Schroeder\IEEEauthorrefmark{2}
    }
    \IEEEauthorblockA{
        \IEEEauthorrefmark{2} IEE S.A., Luxembourg\\ gabriel.beltrao@iee.lu}
}
\maketitle
\begin{abstract}

Continuous monitoring of vital signs is of paramount importance. These critical physiological parameters play a crucial role in the early detection of conditions that affect the well-being of a patient. However, conventional contact-based devices are inappropriate for long-term monitoring. Besides mobility restrictions, they can cause epidermal damage, and even lead to pressure necrosis. In this paper we present a selection of recent works towards enabling the contactless monitoring of vital signs using radar devices. The selected contributions are threefold: an algorithm for recovering the chest wall movements from radar signals; a random body movement and interference mitigation technique; and a robust and accurate adaptive estimation framework. These contributions were tested in different scenarios, spanning ideal simulation settings, real data collected while imitating common working conditions in an office environment, and a complete validation with premature babies in a critical care environment. 

\end{abstract}

\section{Introduction}

Vital signs are a group of biological indicators that show the status of the body’s life-sustaining functions. They provide objective measurements of essential physiological functions and their assessment is the first step for any clinical evaluation. Vital sign information provides valuable insight into the patient's condition, including how they are responding to medical treatment and, more importantly, whether the patient is deteriorating. Abrupt changes in vital signs typically correlate with changes in the cardiopulmonary status and often indicate that a higher level of attention is needed~\cite{Tarassenko2006}.

However, conventional contact-based devices are inappropriate for long-term continuous monitoring. Besides mobility restrictions, they can cause epidermal damage, and even lead to pressure necrosis. On the other hand, radars have already been proven to be a promising technology for contactless monitoring of vital signs. Unlike camera-based systems, radar signals can penetrate through different materials (such as clothing and blankets), and are not affected by skin pigmentation or light levels. Unlike wearable sensors, radar systems do not require users to wear or carry any additional equipment. People under observation do not need to interact with any device, or even to comply with instructions that would change their routines~\cite{Kernec2019}. Radar-based \gls{VSM} has also advantages when considering its applicability in foggy, non-line-of-sight, and through-wall scenarios~\cite{BinObadi2021}. In addition, radar data preserve privacy as no images or videos are recorded~\cite{Zhang2019}.


Despite a few recent works investigating more complex scenarios, most research on contactless \gls{VSM} with radar sensors still focus on a single-person setup under ideal conditions~\cite{Li2019}. The subject is typically instructed to remain relatively motionless (sitting still or lying down), in a quiet environment, and in the absence of other moving objects. The reasons behind this limitation are related to the lack of robustness in dealing with practical scenarios. Particularly, two problems deserve special attention: the additional \glspl{RBM} from the monitored patient, and the harmonic interference from breathing over the heartbeat signal~\cite{statistical}. In addition, algorithm validation is usually still performed in idealized scenarios, where the subject is breathing calmly, and body movements are being emulated through predefined behavior.  

In this paper we present a selection of recent works towards enabling the contactless monitoring of vital signs using radar devices. The selected contributions are threefold: a new algorithm for recovering the chest wall movements from radar signals; a novel random body movement and interference mitigation technique; and a robust and accurate adaptive estimation framework. These contributions were tested in different scenarios, spanning ideal simulation settings, real data collected while imitating common working conditions in an office environment, and a complete validation with premature babies in a critical care environment. 

\section{Methods}

The chest moves during the inspiration/expiration cycle as a result of the diaphragm and intercostal muscle movements. The volumetric changes in the heart muscle due to pumping blood through the circulatory system can also be transmitted to the chest leading to a subtle movement. These small and periodic displacements can be detected by radar, allowing accurate estimation of the breathing and heart rates under certain conditions.

\begin{figure*}[!ht]
\centering
\includegraphics[trim=9cm 7cm 4cm 4cm, clip=true,scale = 0.6]{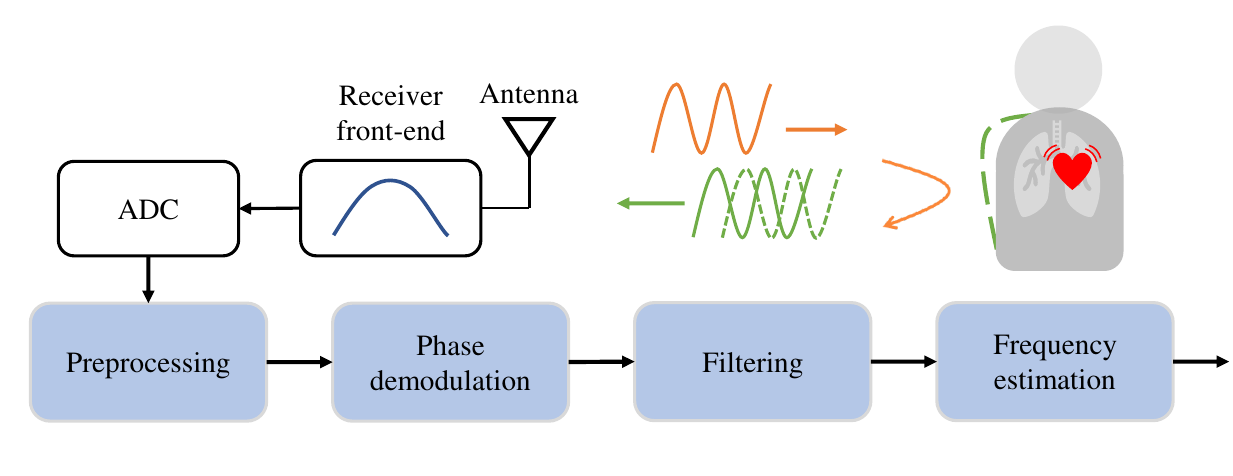}%
\caption{Basic \gls{DSP} block diagram for contactless monitoring of vital signs.}
\label{fig:block_diagram}
\end{figure*}

The desired vital sign information, \textit{i.e.} the chest displacement, is embedded in the phase of the received radar signal. Figure~\ref{fig:block_diagram} shows the basic operating principle and signal processing block diagram. It usually starts with the preprocessing block, which receives samples from the \gls{ADC}, detects the monitored subject, and extracts the corresponding signal modulated by the chest movements. 

The phase demodulation block then aims to recover these movements from the selected signal. As we will show, conventional phase demodulation methods have limitations in practical situations, especially when dealing with interfering \glspl{RBM}. Usually, if the interfering signal is not spectrally overlapped with the vital sign information, it can be easily filtered out by conventional spectral analysis. However, inaccurate phase demodulation may introduce nonlinear errors that can hinder filtering and prevent subsequent frequency estimation even under the simplest interference. To enable robust phase demodulation in practical scenarios, we present an algorithm for recovering the chest wall motion from radar data~\cite{10051975}. It exploits the framework of \gls{US}~\cite{Bhandari2017} to greatly extend the resilience of the recovery process. In this way, robust demodulation can be achieved, even under strong \gls{RBM} interference.

The filtering block is responsible for improving the \gls{SINR} before estimation. The idea is to attenuate the system's noise or any interfering components still present in the demodulated signal. This includes one of the main challenges in monitoring vital signs: how to mitigate the interfering effects of \glspl{RBM}. In practical monitoring situations, the subject may often move body parts like hands, legs, or torso, and even the entire body. The amplitude of these reflected signals is often much stronger than the millimeter-scale chest wall motion, which will potentially be masked by this interference. Since spontaneous \glspl{RBM} are inevitable, solving this problem is fundamental to reliable \gls{VSM} in practical applications. In this direction, we will present our work at the \gls{NICU}~\cite{Beltrao2022a}, where we monitored the respiration of premature babies. The specificity of the monitored patients creates several challenges which were addressed through a novel random body movement mitigation technique, based on the time-frequency decomposition of the recovered signal.

Finally, the aim of the frequency estimation block is to calculate the breathing rate and/or the heart rate. As we mentioned before, higher-order harmonics from the breathing signal may interfere with heart rate estimation and should be considered in the estimator design. To overcome this challenge, we will also present an adaptive estimation framework that explores the harmonic structure existing in the recovered displacement signal~\cite{9965942}. To avoid the harmonic interference from breathing, we generate multiple heart rate estimates based on the heartbeat's own harmonics, and we use a Kalman filter stage to select the most reliable estimates. This avoids low-\gls{SNR}, frequency ambiguous, and harmonic-interfered candidates. In this case, validation was performed with real data collected while imitating common working conditions in an office environment.

\section{Conclusion}

The aforementioned algorithms were able to precisely recover the chest wall motion, effectively reducing the interfering effects of random body movements, and allowing clear identification of different breathing patterns. This capability is the first step toward frequency estimation and early non-invasive diagnosis of cardiorespiratory problems. In addition, most of the time, the adaptive estimation framework provided breathing and heart rate estimates within the predefined error intervals, being capable of tracking the reference values in different scenarios. 

Despite our attempt to move a bit forward in investigating more realistic scenarios, there is still a big gap when considering the practical deployment of a radar-based solution for contactless monitoring of vital signs. Several challenges still need to be addressed, most of them related to the dynamics of monitoring in a practical environment. Particularly, interesting, yet unanswered, research questions arise when monitoring multiple subjects moving freely. Even though target separability, tracking, and range migration are classic radar problems, they become much more complex when considering the simultaneous extraction and processing of vital sign information when the monitored subjects are moving. The longer \glspl{CPI} needed for acquiring at least a few breathing cycles yield to an additional energy trade-off. For instance, if considering a processing window of 12 seconds (only 2 breathing cycles at a breathing rate of 10 bpm), the monitored subject may move almost 10 meters while walking comfortably. This means that the vital sign energy will be spread over dozens of range/angle cells, which can completely jeopardize detection and subsequent estimation.

\bibliographystyle{IEEEtran}
\bibliography{references,myLibrary,uks,extras}

\begin{thebibliography}{10}
\providecommand{\url}[1]{#1}
\csname url@samestyle\endcsname
\providecommand{\newblock}{\relax}
\providecommand{\bibinfo}[2]{#2}
\providecommand{\BIBentrySTDinterwordspacing}{\spaceskip=0pt\relax}
\providecommand{\BIBentryALTinterwordstretchfactor}{4}
\providecommand{\BIBentryALTinterwordspacing}{\spaceskip=\fontdimen2\font plus
\BIBentryALTinterwordstretchfactor\fontdimen3\font minus
  \fontdimen4\font\relax}
\providecommand{\BIBforeignlanguage}[2]{{%
\expandafter\ifx\csname l@#1\endcsname\relax
\typeout{** WARNING: IEEEtran.bst: No hyphenation pattern has been}%
\typeout{** loaded for the language `#1'. Using the pattern for}%
\typeout{** the default language instead.}%
\else
\language=\csname l@#1\endcsname
\fi
#2}}
\providecommand{\BIBdecl}{\relax}
\BIBdecl

\bibitem{Tarassenko2006}
L.~Tarassenko, A.~Hann, and D.~Young, ``{Integrated monitoring and analysis for
  early warning of patient deterioration},'' \emph{Br. J. Anaesth.}, vol.~97,
  no.~1, pp. 64--68, 2006.

\bibitem{Kernec2019}
J.~{Le Kernec} \emph{et~al.}, ``{Radar signal processing for sensing in
  Assisted living},'' \emph{IEEE Signal Process. Mag.}, no. July, pp. 29--41,
  2019.

\bibitem{BinObadi2021}
A.~{Bin Obadi} \emph{et~al.}, ``{A Survey on Vital Signs Detection Using Radar
  Techniques and Processing with FPGA Implementation},'' \emph{IEEE Circuits
  Syst. Mag.}, vol.~21, no.~1, pp. 41--74, 2021.

\bibitem{Zhang2019}
Y.~Zhang, F.~Qi, H.~Lv, F.~Liang, and J.~Wang, ``{Bioradar Technology: Recent
  Research and Advancements},'' \emph{IEEE Microw. Mag.}, vol.~20, no.~8, pp.
  58--73, aug 2019.

\bibitem{Li2019}
C.~Li, ``{Vital-sign monitoring on the go},'' \emph{Nat. Electron.}, vol.~2,
  no.~6, pp. 219--220, 2019.

\bibitem{statistical}
G.~Beltr{\~a}o, M.~Alaee-Kerahroodi, U.~Schroeder, D.~Tatarinov, and M.~R.
  Bhavani~Shankar, ``Statistical performance analysis of radar-based
  vital-sign processing techniques,'' in \emph{Sensing Technology}.\hskip 1em
  plus 0.5em minus 0.4em\relax Springer International Publishing, 2022, pp.
  101--112.

\bibitem{10051975}
G.~Beltrão, T.~Feuillen, M.~B. Shankar, M.~Alaee-Kerahroodi, and U.~Schroeder,
  ``Unlimited sampling for radar-based vital sign monitoring,'' in \emph{2022
  56th Asilomar Conf. on Signals, Systems, and Comp.}, 2022, pp. 1131--1136.

\bibitem{Bhandari2017}
A.~Bhandari, F.~Krahmer, and R.~Raskar, ``{On unlimited sampling},'' \emph{2017
  12th Int. Conf. Sampl. Theory Appl. SampTA 2017}, pp. 31--35, 2017.

\bibitem{Beltrao2022a}
G.~Beltr{\~{a}}o \emph{et~al.}, ``{Contactless radar-based breathing monitoring
  of premature infants in the neonatal intensive care unit},'' \emph{Sci.
  Rep.}, vol.~12, no.~1, pp. 1--15, 2022.

\bibitem{9965942}
G.~Beltrão, W.~A. Martins, B.~Shankar M.~R., M.~Alaee-Kerahroodi,
  U.~Schroeder, and D.~Tatarinov, ``Adaptive nonlinear least squares framework
  for contactless vital sign monitoring,'' \emph{IEEE Transactions on Microwave
  Theory and Techniques}, vol.~71, no.~4, pp. 1696--1710, 2023.

\end{thebibliography}
\end{document}